\documentclass[manuscript]{aastex}

\newcommand{\be}{\begin{equation}}
\newcommand{\ee}{\end{equation}}
\newcommand{\bea}{\begin{eqnarray}}
\newcommand{\eea}{\end{eqnarray}}

\begin{document}

\title{ A Better Way to Reconstruct Dark Energy Models ?}
\author {A. Sil\altaffilmark{1}}
\email{amitavadrsil@rediffmail.com}
\and
\author{S.Som\altaffilmark{2}}
\email{sumitsom79@yahoo.com}
\affil{Relativity and Cosmology Research Centre,\\Department of Physics, Jadavpur University,\\
Kolkata - 700032, India.} 
\altaffiltext{1}{St.Paul's C. M. College, 33/1 Raja Rammohan Sarani, Kolkata 700009, India.}

\altaffiltext{2}{Meghnad Saha Institute of Technology, Madurdaha, Uchhepota, Kolkata-700150, India.}

\begin{abstract}
To reconstruct dark energy models the redshift $z_{eq}$, marking the end of radiation era and the beginning of matter-dominated era, can play a role as important as $z_{t}$, the redshift at which deceleration parameter experiences a signature flip. To implement the idea we propose a variable equation of state for matter that can bring a smooth transition from radiation to matter-dominated era in a single model. A popular $\Lambda \propto \rho$ dark energy model is chosen for demonstration but found to be unacceptable. An  alternative $\Lambda \propto \rho a^{3}$ model is proposed and found to be more close to observation.
\end{abstract}


\section{Introduction}
The current trend in modern cosmology is mainly focussed on issues of possible explanations of the observed late time acceleration in the expansion of our Universe\citep{riess,perl}. The presence of a dark energy of unknown origin which dominates over matter in our Universe is now almost accepted. Unlike ordinary matter the dark energy has a repulsive effect and its dominance accounts for the observed acceleration. Several candidates of dark energy are found in literature. Among them the most elementary is the cosmological constant $\Lambda$\citep{padd,peebles}. Variable $\Lambda$ models\citep{car,wet,arbab,paddy,vis,shap,dutta} were introduced later in an effort to overcome the fine tuning problem. Among other candidates most popular are different kinds of scalar fields like quintessence\citep{caldwell1}, K-essence\citep{chiba}and phantom fields\citep{caldwell2} to name only a few. The Chaplygin gas\citep{kamen} has an equation of state that can produce negative pressure in later phase of evolution to account for the accelerated expansion of the Universe. Some geometrical origin of the late-time acceleration are also found in modified gravity models\citep{nojiri}. All these models have their own merits and demerits. The exact nature of dark energy is therefore yet to be ascertained. For a detailed review on different dark energy candidates one may see the work of \citet{sahni} or \citet{copeland}.

Most of these models are usually studied only in the matter dominated phase assuming the presence of some kind of dark energy to explain the transition from deceleration to acceleration. This is due to the fact that radiation era was very short lived and dominated the Universe in its early phase of evolution whereas acceleration is a more recent phenomena which has occurred very late in the matter dominated epoch. The viability of such models are usually checked by comparing the variation of deceleration parameter $q$ with the red-shift $z$ obtained theoretically with that from observations. A good estimation of $z=z_t$, the redshift marking the transition from deceleration to acceleration, is available from observation. But observational data are limited over a small range of $z$ values around $z_t$. Hence many models that fit observational data are found to behave quite differently outside this range. To choose among these models we need at least another event that happens to be outside this range of $z$, preferably in the distant past. We point out that a very good estimate of $z=z_{eq}$, the redshift when matter density was equal to radiation density, is at our disposal. Thus instead of a single point we now have two well separated points through which the $q$ vs. $z$ curve for any viable cosmological model should pass. Such a model would be more reliable as its viability could be ensured over a large range of $z$ values. But then to estimate $z_{eq}$ we must have both radiation and matter in a single model.
\par In the next section we propose a variable equation of state for matter that mimics radiation in the past and pressureless dust in the later course of evolution. In section 3 we examine the $\Lambda \propto \rho$ model with the proposed equation of state for matter. In section 4 an alternative variable $\Lambda $ model is proposed.

\section{Radiation and matter in a single model}

To begin with let us keep aside dark energy for a while and follow the standard practice of idealizing the Universe as homogeneous and isotropic supported by the isotropy of microwave background radiation. The spatial geometry of the Robertson-Walker spacetime is chosen flat in view of strong observational evidences\citep{zhao}. In these kinds of standard models of the Universe with no dark-energy, radiation dominates for roughly the first 2000 years of evolution and matter dominates there-after\citep{peacock}. In Friedmann models the homogeneous and isotropic Universe is filled with an ideal fluid. The equation of state for this fluid is either that of radiation ($p_{r}=\frac{1}{3}\rho_{r}$) or that of pressureless dust ($p_{m}=0$). Thus one needs two different cosmological models with different equations of state for matter to explain the expansion history in two different epochs of the same Universe. No single Friedmann model can describe the full history of the Universe. The deceleration parameter $q$ also takes different positive values, however constant, in each epoch. Such cosmological models are usually studied by joining a radiation Universe ($p=\rho /3$) to a matter dominated Universe 
($p=0$), smoothly at some particular time. 

\par Attempts were made to describe the entire evolution in a single model by \citet{chernin1,chernin2}, \citet{mcintosh1,mcintosh2}, \citet{lb}, \citet{jacobs} and \citet{cohen} in late sixties of the last century. In those models the homogeneous and isotropic Universe is assumed to be filled with a perfect fluid which has two components {\it viz.} radiation and matter. Assuming no interaction between radiation and matter, the radiation energy density $\rho_{r}$ decreases with time in the usual manner $\rho_{r} \sim a^{-4}$ while the matter density falls as $\rho_{m} \sim a^{-3}$. Thus the total energy density $\rho$ and the total pressure $p$ of the fluid may be written as 
\bea
\rho = \rho_{r} + \rho_{m} = \rho_{r0}(a_{0}/a)^{4}+\rho_{m0}(a_{0}/a)^{3}\\
p = p_{r}+p_{m}=\frac{1}{3}\rho_{r0}(a_{0}/a)^{4}
\eea
where the subscript `$0$' indicates the value at present epoch. We thus find that $\rho$ and $p$ maintain an equation of state
\be
f=\frac{p}{\rho}=\frac{1}{3}\left[ 1+( \rho_{m0}/\rho_{r0})(a/a_{0})\right]^{-1}=\frac{1}{3}\left[ 1+\frac{a}{a_{eq}}\right]^{-1}  \label{eos1}
\ee
where $a_{eq}$ is defined as the scale when $\rho_{r}$ and $\rho_{m}$ are equal {\it i.e.}
\be
\rho_{r0}(a_{0}/a_{eq})^{4}=\rho_{m0}(a_{0}/a_{eq})^{3}.
\ee
One may see that for small values of `$a$' the equation of state approximates to that of pure radiation while for $a>>a_{eq}$ the pressure becomes negligible and mimics the matter dominated epoch.

\par We propose the cosmological fluid to obey the above equation of state. With such a model in presence of dark energy it becomes possible and easier to tune parameters of the model in order to obtain desired or observed values for both $z_{eq}$ and $z_{t}$ simultaneously. As an exercise, in the next section, we shall consider a variable $\Lambda$ `dark energy' model to demonstrate this point.

\section{Testing a $\Lambda \propto \rho$ model}

We need an example of a dark energy model to show how our idea can be implemented. In absence of any knowledge about the exact nature of dark energy majority of approaches are phenomenological where we limit the equation of state for dark energy from observational data. Although observation does not rule out the possibility of a varying equation of state for dark energy\citep{kujat,bartel,knop,tegmark} the evidences are strongly in favour of an equation of state that is very close to $p_{DE}/\rho_{DE}=-1$\citep{davis}. Thus observational data supports the choice of Einstein's cosmological constant $\Lambda$ as a dark energy. However, a major challenge in constant $\Lambda$ cosmology is perhaps to explain its vanishingly small value at present. From field theoretical point of view if $\Lambda$ represents the vacuum energy density then its value should be $\sim 10^{123}$ times its observed value today\citep{weinberg}. Attempts were made to explain such anomalies in the order of magnitude of $\Lambda$ by introducing the idea of a time varying $\Lambda$. 

\par There exists a large number of variable $\Lambda$ models in the literature\citep{car,wet,arbab,paddy,vis,shap,dutta,ray1}. A popular choice for variable $\Lambda$ in the literature is a function that either decays like the energy density $\rho$ or like the square of the Hubble parameter or as $\ddot{a}/a$. However, all these models were found to be equivalent\citep{ray}. A difficulty in such models is to find a `single' positive constant of proportionality that can produce deceleration in the radiation dominated era as well as a transition to acceleration late in the matter dominated era. This is indeed obvious (though undesirable) as in such models the variations of the matter density $\rho$ are distinctly different in radiation and matter dominated eras. Thus with a variable equation of state for matter that smoothly connects the radiation and matter era in a single model, it might be interesting to estimate the value of the constant of proportionality, that produces acceptable values for observational parameters. In this section we choose a variable $\Lambda \propto \rho$ as the dark energy candidate and investigate how far it can produce observational results including a desired value of $z_{eq}$ when we introduce a variable equation of state for matter. 
Keeping in mind the above said difficulty  we will relax the reasoning behind the choice of the equation of state for matter  by introducing a parameter $m$ in the expression for $f$ in equation (\ref{eos1}) preserving its asymptotic behavior
\be
f=\frac{1}{3}\left[1+\left(\frac{a}{a_{eq}}\right)^{m}\right]^{-1} \label{eos2}
\ee
where $m$ is a positive constant that may be determined from observational constraints. 

\par With a choice of the energy-momentum tensor of the Universe 
\be
\tilde{T}^{\mu\nu}=T^{\mu\nu}-\frac{c^4\Lambda}{8{\pi}G}g^{\mu\nu}=(\rho+p)u^{\mu}u^{\nu}-pg^{\mu\nu}-\frac{c^4\Lambda}{8{\pi}G}g^{\mu\nu}
\ee
we write the Einstein's field equations 
\be 
R^{\mu\nu}- \frac{1}{2}Rg^{\mu\nu}=\frac{-8\pi G}{c^4}\tilde{T}^{\mu\nu} \label{ee}
\ee 
for a homogeneous isotropic Robertson-Walker type spacetime with flat ($k=0$) spatial geometry as
\bea
\frac{\dot{a}^2}{a^2} = 2u\rho + \frac{1}{3}c^2\Lambda \label{fe1}\\
\frac{\ddot{a}}{a} = -u(\rho+3p)+\frac{1}{3}c^2\Lambda \label{fe2}
\eea
where $u=\frac{4\pi G}{3c^2}$. 
\par From equation (\ref{fe1}) it may appear that with $\Lambda \propto \rho$, one can trivially absorb $\Lambda$ in to the constant $u$ and hence there would be no dark energy in the cosmological evolution. This is indeed true for a perfect fluid with an equation of state $p=\kappa\rho$. The introduction of $\Lambda \propto \rho$ effectively modifies the proportionality constant $\kappa$ and the model becomes equivalent to one with a new fluid with no dark energy. But with a choice of variable equation of state for matter one does not have the same liberty in equation (\ref{fe2}). 
Following Bianchi identity, the divergence of the field equation (\ref{ee}) gives,
\be
[R^{\mu\nu}- \frac{1}{2}Rg^{\mu\nu}]_{;\nu} = 0 =\frac{-8\pi G}{c^4} [T^{\mu\nu}-\frac{c^4\Lambda(t)}{8\pi G}g^{\mu\nu}]_{;\nu}
\ee 
which indicates that empty spacetime cannot be obtained as solutions of the field equations if $\Lambda$ varies with time. Thus a variable $\Lambda$ is consistent with Mach's principle that for a meaningful spacetime geometry presence of matter is a necessity\citep{vis}. The conservation of energy-momentum tensor requires 
\be
u\dot{\rho}+\frac{1}{6}\dot{\Lambda}c^2+3u(\rho+p )\frac{\dot{a}}{a}=0. \label{conservation}
\ee
The above equation clearly shows the presence of interaction between matter-radiation and dark energy. 

\par Among the set of three equations (\ref{fe1}), (\ref{fe2}) and (\ref{conservation}) only two are independent, whereas number of unknown functions are four. In order to get an exact solution to the problem we make the following choices. We make an ansatz that the effective equation of state for the matter-radiation part as given in equation (\ref{eos2}) is

$$f=\frac{1}{3}\left[1+\left(\frac{a}{a_{eq}}\right)^{m}\right]^{-1}$$

and choose the variable cosmological parameter as
\be
\Lambda = \frac{u}{c^2}\lambda \rho.  \label{lambda}
\ee
Here $\lambda$ is a positive constant. For the above choices equations (\ref{fe1}) and (\ref{fe2}) reduce to
\bea
\frac{\dot{a}^2}{a^2}=u\rho \left(2+\frac{\lambda}{3}\right)\\
\frac{\ddot{a}}{a}=-u\rho \left(1+3f-\frac{\lambda}{3}\right).
\eea
The deceleration parameter $q$ is therefore given by
\be
q=-\frac{\ddot{a}/a}{\dot{a}^{2}/a^{2}}=\frac{1+3f-\frac{\lambda}{3}}{2+\frac{\lambda}{3}}. \label{q}
\ee
Choice of $f$ as in equation (\ref{eos2}) implies that for small values of $a$ the deceleration parameter behaves as $\frac{2-\frac{\lambda}{3}}{2+\frac{\lambda}{3}}$. If we believe that in early radiation dominated era the Universe went through a decelerated phase with $q\sim 1$ then $\lambda$ must be vanishingly small in the radiation era. In the later phase for $a>>a_{eq}$, the function $f$ becomes very small and $q$ behaves as $\frac{1-\frac{\lambda}{3}}{2+\frac{\lambda}{3}}$. Hence to observe late-time acceleration in the expansion we must have $\lambda >3$, preferably $\lambda \geq 12$ so that $q\leq -0.5$ for large $a$ to make it consistent with observation. Restriction on $\lambda$ also comes from the observation of density parameters $\Omega_{m}$ and $\Omega_{\Lambda}$ defined as 
\bea
\Omega_{m}=\frac{2u\rho}{\dot{a}^{2}/a^{2}}=\frac{6}{6+\lambda}\\
\Omega_{\Lambda}=\frac{c^{2}\Lambda}{3\dot{a}^{2}/a^{2}}=\frac{u\lambda\rho}{3\dot{a}^{2}/a^{2}}=\frac{\lambda}{6+\lambda}.
\eea
Observed values of $\Omega_{m}$ and $\Omega_{\Lambda}$ at the present epoch are $\Omega_{m0}\approx 0.3$ and $\Omega_{\Lambda0}\approx 0.7$\citep{rs}. Such observation puts a limit on the present value of $\lambda$ as $\lambda_0\approx 14$.

\par With the expression for $f$ as in equation (\ref{eos2}) the expression for deceleration parameter $q$ in view of equation (\ref{q}) becomes
\be
q=\frac{3[(\frac{a}{a_{eq}})^{m}+1]^{-1}+(3-\lambda)}{6+\lambda}. \label{q3}
\ee
It is evident from the above equation that to ensure a decelerated phase in the beginning of evolution and a late time transition from decelerated to accelerated phase in the expansion of the Universe we require $3<\lambda<6$. If the value of $\lambda$ lies within this range one can see that the range for $q$ will be $-\frac{1}{4} < q <\frac{1}{3}$. We can see that this makes it difficult to produce observational requirements. The radiation-matter density $\rho$ can be solved analytically from conservation equation (\ref{conservation}) to get
\be
\rho = A[a^{-4}(a^m+a_{eq}^m)^{\frac{1}{m}}]^{\frac{6}{\lambda+6}} \label{rho3}
\ee
where $A$ is the constant of integration. Notice that for $\lambda =0$ (i.e no $\Lambda$ parameter) and for any value of $m>0$ the energy density $\rho \sim a^{-4}$ for small values of $a$ (such that $a << a_{eq}$) whereas for large values of $a$ (such that $a>>a_{eq}$), $\rho \sim a^{-3}$. Substituting $(\ref{rho3})$ back in $(\ref{fe1})$ we can solve for $a$ for different possible values of $m$. For a choice, $m=\frac{12}{(\lambda +6)}$ the equation can be easily solved to obtain
\be
a=[(\alpha t + a_{eq}^{\frac{3m}{4}})^{\frac{4}{3}}-a_{eq}^m]^{\frac{1}{m}}\label{scale3}
\ee
where, $\alpha$ is a constant involving $\lambda $ and $A$
\be
\alpha = \frac{9}{(\lambda +6)}\left[Au\left(2+\frac{\lambda}{3}\right)\right]^\frac{1}{2}.
\ee
Note that the particular choice $m = \frac{12}{(\lambda +6)}$ gives $m=2$ for $\lambda =0$ (i.e no $\Lambda$ parameter) and hence from equation $(\ref{scale3})$ we get
\be
a=\left[ a_{eq}^2\left(1+\frac{\alpha t}{a_{eq}^{\frac{3}{2}}}\right)^{\frac{4}{3}}-a_{eq}^2\right]^{\frac{1}{2}}.
\ee
For such a model with no $\Lambda $ parameter the scale factor $a \sim  t^{\frac{1}{2}}$ for small values of $t$ (radiation dominated era), and for large values of $t$ (matter dominated era), $a \sim t^{\frac{2}{3}}$. Thus instead of two different (radiation dominated and matter dominated FRW) models we can have a single model with variable equation of state that has the desirable asymptotic behavior. This finding may be identified with those of \citet{jacobs}.

\par It may be noted at this point that in general there exists no relation between $m$ and $\lambda$. If we do not go for any particular choice of $m$ it might be interesting to see how the observational constraints can relate them. If $a=a_t$ marks the transition from deceleration to acceleration in the expansion of the Universe we can see from equation (\ref{q3}) that 
\be
\left[\frac{a_t}{a_{eq}}\right]^m = \frac{6- \lambda}{\lambda -3}.
\ee
Since the acceleration of the Universe has occured late in the matter dominated era, the above ratio must be greater than 1 and hence $\lambda <4.5$. This resets the limits on $\lambda$ as $3<\lambda <4.5$.
\citet{rs} and \citet{tu} have shown the redshift-distance relation of SN Ia compared to a family of flat, $\Omega_{\Lambda}$ cosmologies where the transition between accelerating and decelerating epochs occurs at a redshift
\[ z_{t}=\left(\frac{2\Omega_{\Lambda}}{\Omega_{m}}\right)^{\frac{1}{3}}-1. \]
For $\Omega_{\Lambda}\approx  0.7$ and $\Omega_{m}\approx  0.3$, the above relation gives 
\[ 1+z_{t}\equiv \frac{a_0}{a_{t}}\approx 1.7. \]
Also the redshift $z_{eq}$ marking the beginning of matter dominance over radiation is usually given by\citep{peacock}
\[ 1+z_{eq}\equiv \frac{a_0}{a_{eq}}\approx 1000. \]
 
\begin{figure}[h]
\epsscale{0.50}
\plotone{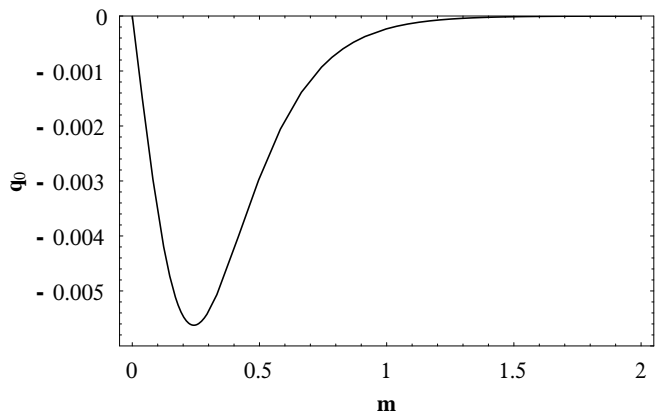}
\caption{\footnotesize $q_0$ vs $m$ plot. The minima in the graph shows that $q_0$ can never be made smaller than $-5.623\times10^3$.}
\label{fig1}
\end{figure}

\begin{figure}[h]
\plotone{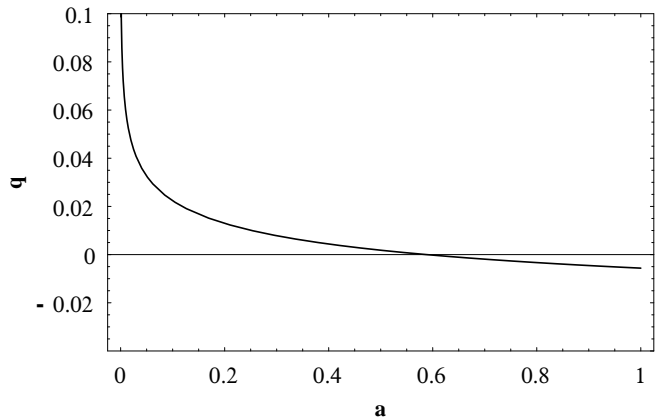}
\caption{\footnotesize $q$ vs $a$ plot. The plot shows the variation of $q$ with $a$ for $m=0.24207$ and $\lambda=3.528$.}
\label{fig2}
\end{figure}

Assuming $\frac{a_t}{a_{eq}} \approx \frac{10^3}{1.7}$, we find that $m$ and $\lambda$ obey the relation
\be
\left(\frac{10^3}{1.7}\right)^m=\frac{6-\lambda}{\lambda-3}.\label{ml}
\ee
We use equation (\ref{ml}) to eliminate $\lambda$ from equation (\ref{q3}). The present day value of the deceleration parameter $q_0$ is now a function of $m$ as given below
\be
q_0=\frac{10^{3m}\left(\frac{1}{1.7^m}-1\right)}{(1+10^{3m})(4+3(\frac{10}{1.7})^{3m})}. \label{qnot}
\ee
When we plot the above function in fig.(\ref{fig1}) we find there exists a minima at $m\approx 0.242$ (and hence $\lambda\approx 3.528$). The minimum value of $q_0$ is however very small ($\approx -5.623\times 10^{-3}$). There exists no other choice of the pair $m$ and $\lambda$ for which $q_0$ could be made smaller than this value. Thus we find that a constant $\lambda$ that fits both the radiation and matter dominated models is capable of producing a signature flip in the deceleration parameter. However within the framework of our model the value of $q_0$ becomes insignificantly small due to observational constraints on the values of $a_{eq}$ and $a_t$. The variation of $q$ with $a$ is plotted in fig.(\ref{fig2}) in $a_0=1$ unit for a pair of $m$ and $\lambda$ that produces maximum possible acceleration at the present epoch in this model.\\

\section{An alternative variable $\Lambda $ model that passes the test}

In view of the failure of a variable $\Lambda$ model of the kind $\Lambda\propto\rho$ to meet observational constraints as well as desired features as shown in the previous section we propose that 
a simple choice for variable $\Lambda$ satisfying the observational constraints might be 
\be
\Lambda =\frac{u}{c^{2}}\lambda(a)\rho  \label{2}
\ee
where
\be
\lambda(a) = 3\left(\frac{a}{a_t}\right)^{3}. \label{1}
\ee
The motivation behind such a choice is the success of 'Lambda Cold Dark Matter' ($\Lambda$CDM) model\citep{padd} in explaining the observed acceleration. Anticipating that in the matter dominated era the energy density $\rho$ would vary as $a^{-3}$, we propose (\ref{2}) so that $\Lambda$ may behave as a cosmological constant in the later phase of evolution. This way the field theoretical demand of very large value of $\Lambda$ in the early Universe will also be met if $\rho \sim a^{-4}$ for very small values of $a$.
\par
\begin{figure}[h]
\plotone{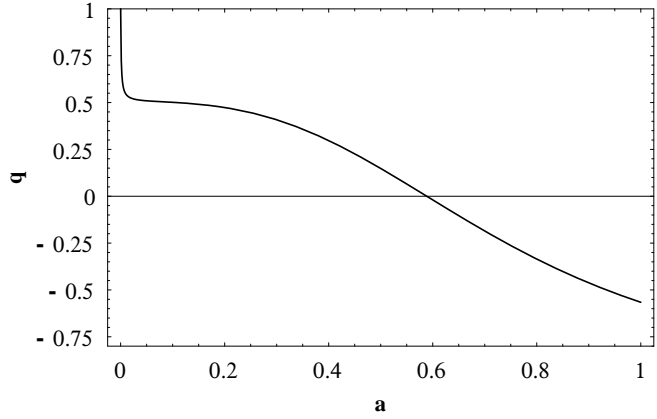}
\caption{\footnotesize{$q$ vs $a$ plot.}}
\label{fig3}
\end{figure}

 
Using equation (\ref{1}) and equation (\ref{eos1}) the deceleration parameter $q$, given by equation (\ref{q}), in this model takes up the following functional form
\be
q=\frac{\left(1+\frac{a}{a_{eq}}\right)^{-1}+3}{2+\left(\frac{a}{a_t}\right)^3}-1.
\ee
The asymptotic behavior of $q$ shows that it varies smoothly from $+1$ in distant past to $-1$ in distant future with its signature flip occuring at $a\approx a_t$. When $q$ is plotted against $a$ ($a_0=1$ unit) in fig.(\ref{fig3}) we find that for a considerable period of time in the early stage of matter dominated era $q$ remains almost constant at a value $+0.5$. This is indeed a very desirable feature of this model as it provides a favourable condition for structure formation. A rough estimation with $z_{eq}\approx 1000$ and $z_t\approx 0.7$ shows that the present value of deceleration parameter $q_0\approx -0.56$.\\

\par
\begin{figure}[h]
\plotone{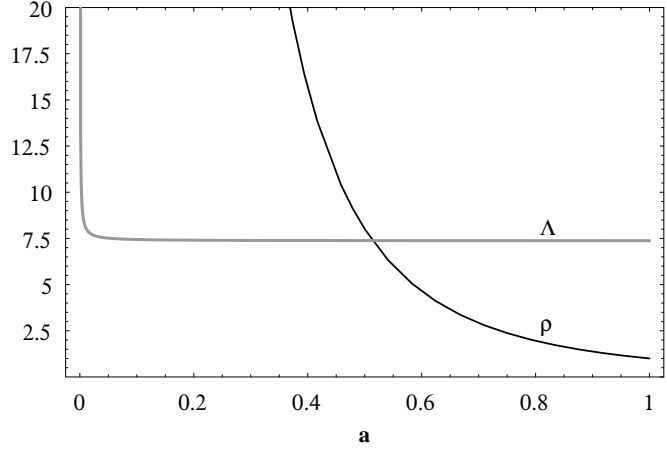}
\caption{\footnotesize $\rho$ , $\Lambda$ vs $a$ plot. In the early stage of evolution $\Lambda \sim a^{-1}$ as $\rho \sim a^{-4}$. $\Lambda$ gradually attains a constant value with evolution since $\rho \sim a^{-3}$ for large $a$.}
\label{fig4}
\end{figure}

\begin{figure}[h]
\plotone{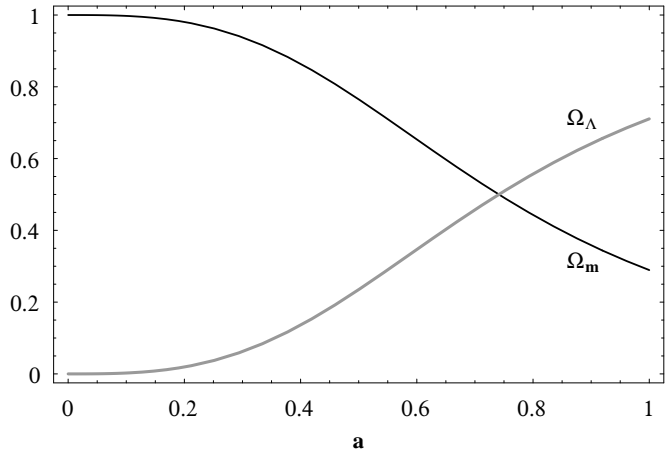}
\caption{\footnotesize{$\Omega_{m}$ , $\Omega_{\Lambda}$ vs $a$ plot.}}
\label{fig5}
\end{figure}

 The conservation equation (\ref{conservation}) with the above choice of $\Lambda$ reduces to
\be
\frac{\dot{\rho}}{\rho}+\left[\frac{3\lambda}{6+\lambda}+2(q+1)\right]\frac{\dot{a}}{a}=0 \label{conservation1}
\ee
which implies$$\ln\rho=\ln{A}+\ln{a^{-4}}+\ln{(a_{eq}+a)^{2\alpha}}+\ln{(2a_t^3+a^3)^{-2\beta}}+\ln{(2^{2/3}a_t^2-2^{1/3}aa_t+a^2)^{2\gamma}}$$
\be
-\ln{(2^{1/3}+a)^{4\gamma}}+\eta\arctan{\left(\frac{2a}{2^{1/3}\sqrt{3}}-\frac{1}{\sqrt{3}}\right)}\label{rho}
\ee
where $A$ is the arbitrary constant of integration and $\alpha$, $\beta$, $\gamma$ and $\eta$ are constants involving $a_t$ and $a_{eq}$. For $a_{eq}=0.001$ and $a_t=0.59$ in $a_0=1$ unit, the values of the constants are $\alpha=0.5$, $\beta=4.06\times 10^{-10}$, $\gamma=1.12\times 10^{-4}$ and $\eta=7.76\times 10^{-4}$. Thus only the second and the third term on the right hand side of equation (\ref{rho}) contribute significantly to the variation of $\rho$ with $a$. Variation of $\rho$ and $\Lambda$ with $a$ is shown below [fig.(\ref{fig4})] in arbitrary units. The density parameters are also plotted in fig.(\ref{fig5}). In both the figures $a$ is scaled in $a_0=1$ unit.\\
\par The asymptotic behavior of $\rho$ may be easily investigated from equation (\ref{conservation1}). For very small values of $a$ we find that $\lambda\rightarrow 0$ while $q\rightarrow 1$. This makes the conservation equation (\ref{conservation1}) to reduce to
\be
\frac{\dot{\rho}}{\rho}+4\frac{\dot{a}}{a}=0.
\ee
Hence the matter-radiation density $\rho \sim a^{-4}$ in the early phase of evolution whereas for $a\rightarrow \infty$ the functions $\lambda\rightarrow \infty$ while $q\rightarrow -1$ making equation (\ref{conservation1}) to reduce to
\be
\frac{\dot{\rho}}{\rho}+3\frac{\dot{a}}{a}=0.
\ee
Thus in the late stage of evolution the matter-radiation density $\rho \sim a^{-3}$.
This makes $\Lambda$ to attain a constant value in the later course of evolution. \\

\section{Conclusion}

We have shown in this present article that the viability of a dark energy model can be cheked in a better way by demanding that $z_{eq}$ should be $\sim 10^{3}$. Introducing a variable equation of state for matter we have investigated the nature of expansion of the Universe in two variable dark energy models. The variable equation of state for matter ensures that the early epoch of the Universe is dominated by radiation. With the expansion of the Universe the pressure falls as $a^{-m}$ $(m>0)$ to produce an almost pressure-free late-Universe. The equation of state for dark energy is essentially chosen to be $\alpha_{DE}=p_{DE}/\rho_{DE}=-1$ while the $\rho_{DE}$ is in general a function of the scale factor $a$. This in other way amounts to the same as the presence of a dynamical $\Lambda$ in the field equations. It makes the conservation equation notably different. One has to allow interaction among the matter part and dark energy. The functional form of $\Lambda$ is then chosen to produce late-time acceleration in the expansion of the Universe. A $\Lambda \propto \rho$ model fails to meet observational constraints. A choice of $\Lambda \propto \rho a^{3}$ is found to be a better alternative. Thus we conclude that meeting the additional $z_{eq}\sim 10^{3}$ requirement may be an essential feature of reconstructing dark energy models.

\end{document}